\documentclass[aps,pre,amsmath,twocolumn,superscriptaddress]{revtex4-1}
\usepackage[dvipdfmx]{graphicx}
\usepackage{color}
\usepackage{bm}
\usepackage{braket}
\usepackage{amsmath,amssymb}

\begin{document}
\title{Machine learning potentials for multicomponent systems: 
The Ti-Al binary system}
\author{Atsuto \surname{Seko}}
\email{seko@cms.mtl.kyoto-u.ac.jp}
\affiliation{Department of Materials Science and Engineering, Kyoto University, Kyoto 606-8501, Japan}

\date{\today}

\begin{abstract}
Machine learning potentials (MLPs) are becoming powerful tools for performing accurate atomistic simulations and crystal structure optimizations. 
An approach to developing MLPs employs a systematic set of polynomial invariants including high-order ones to represent the neighboring atomic density.
In this study, a formulation of the polynomial invariants is extended to the case of multicomponent systems.
The extended formulation is more complex than the formulation for elemental systems.
This study also shows its application to Ti-Al binary system.
As a result, an MLP with the lowest error and MLPs with high computational cost performance are selected from the many MLPs developed systematically.
The predictive powers of the developed MLPs for many properties, such as the formation energy, elastic constants, thermodynamic properties, and mechanical properties, are examined. 
The MLPs exhibit high predictive power for the properties in a wide variety of ordered structures.
The present scheme should be systematically applicable to other multicomponent systems.
\end{abstract}

\maketitle
\section{Introduction}

Machine learning potentials (MLPs) have been developed from extensive datasets generated by density functional theory (DFT) calculation, and can significantly improve the accuracy and transferability of interatomic potentials.
Therefore, MLPs are becoming useful tools for performing crystal structure optimizations and accurate large-scale atomistic simulations, which are prohibitively expensive by DFT calculation.
Over the last decade, a number of methods that can be used to develop MLPs and their applications have been reported
\cite{
Lorenz2004210,
behler2007generalized,
bartok2010gaussian,
behler2011atom,
han2017deep,
258c531ae5de4f5699e2eec2de51c84f,
PhysRevB.96.014112,
PhysRevB.90.104108,
PhysRevX.8.041048,
PhysRevLett.114.096405,
PhysRevB.95.214302,
PhysRevB.90.024101,
PhysRevB.92.054113,
PhysRevMaterials.1.063801,
Thompson2015316,
wood2018extending,
PhysRevMaterials.1.043603,
doi-10.1137-15M1054183,
PhysRevLett.120.156001,
podryabinkin2018accelerating,
GUBAEV2019148,
doi:10.1063/1.5126336}.
In these studies, the contribution of an atom to potential energy is given as a function of quantities depending on its neighboring environment, called structural features.
Also, several models are employed to describe a mapping from structural features to the atomic contribution, 
including artificial neural network models
\cite{
Lorenz2004210,
behler2007generalized,
behler2011atom,
han2017deep,
258c531ae5de4f5699e2eec2de51c84f,
PhysRevB.96.014112}, 
Gaussian process models
\cite{
bartok2010gaussian,
PhysRevB.90.104108,
PhysRevX.8.041048,
PhysRevLett.114.096405,
PhysRevB.95.214302},
and linear models
\cite{
PhysRevB.90.024101,
PhysRevB.92.054113,
PhysRevMaterials.1.063801,
Thompson2015316,
wood2018extending,
PhysRevMaterials.1.043603,
doi-10.1137-15M1054183}.

Structural features play an essential role in controlling the accuracy and computational efficiency of MLPs, which are conflicting properties in general \cite{PhysRevB.99.214108,hernandez2019fast,doi:10.1021/acs.jpca.9b08723}.
A systematic set of structural features is composed of polynomial invariants.
The polynomial invariants include second- and third-order bond-orientational order parameters \cite{PhysRevB.28.784}, angular Fourier series \cite{bartok2013representing}, the bispectrum \cite{kondor2008group,bartok2013representing}, and moment tensors \cite{doi-10.1137-15M1054183,flusser2009moments}, which have been adopted to develop MLPs and machine learning models of physical properties in compounds.
Recently, a group-theoretical procedure for enumerating the polynomial invariants derived from spherical harmonics, including high-order ones, was proposed \cite{PhysRevB.99.214108}.
The angular Fourier series, bond-orientational order parameters, and bispectrum can be included in the enumeration by this procedure.
MLPs developed with the polynomial invariants for a wide variety of elemental systems exhibit high predictive power for a wide range of structures. 
They are available in the \textsc{Machine Learning Potential Repository} \cite{MachineLearningPotentialRepositoryArxiv,MachineLearningPotentialRepository}.
However, the polynomial invariants and related potential energy models must be generalized for developing MLPs in multicomponent systems.

In this study, polynomial invariants for developing MLPs for a multicomponent system are formulated.
The present formulation of the polynomial invariants should be helpful in developing MLPs for multicomponent systems even within other frameworks.
Polynomial models combined with the polynomial invariants are also introduced to describe the potential energy.
This study also shows an application of the polynomial models to the development of Pareto optimal MLPs in Ti-Al binary alloy system.
The predictive power of the Pareto optimal MLPs is examined for the cohesive energy, the formation energy, the elastic constants, the phonon density of states and dispersion curves, the thermal expansion, the energy profile along the Bain path, and the stacking fault properties.

Section \ref{mlip-TiAl:Sec-Model} introduces potential energy models in multicomponent systems, including the polynomial invariants representing the neighboring atomic density and polynomial models for the potential energy.
In Sec. \ref{mlip-TiAl:Sec-Datasets}, datasets required to develop MLPs for the Ti-Al binary system are explained.
Computational procedures for constructing datasets and estimating coefficients in the potential energy models are also shown.
In Sec. \ref{mlip-TiAl:Sec-Results}, the development of Pareto optimal MLPs for the Ti-Al binary system is demonstrated.
The predictive power of the MLPs for many properties, such as the formation energy, the elastic constants, the thermodynamic properties, and the mechanical properties, is also investigated.
Finally, this paper is concluded in Sec. \ref{mlip-TiAl:Sec-Conclusion}.

\begin{figure}[tbp]
\includegraphics[clip,width=\linewidth]{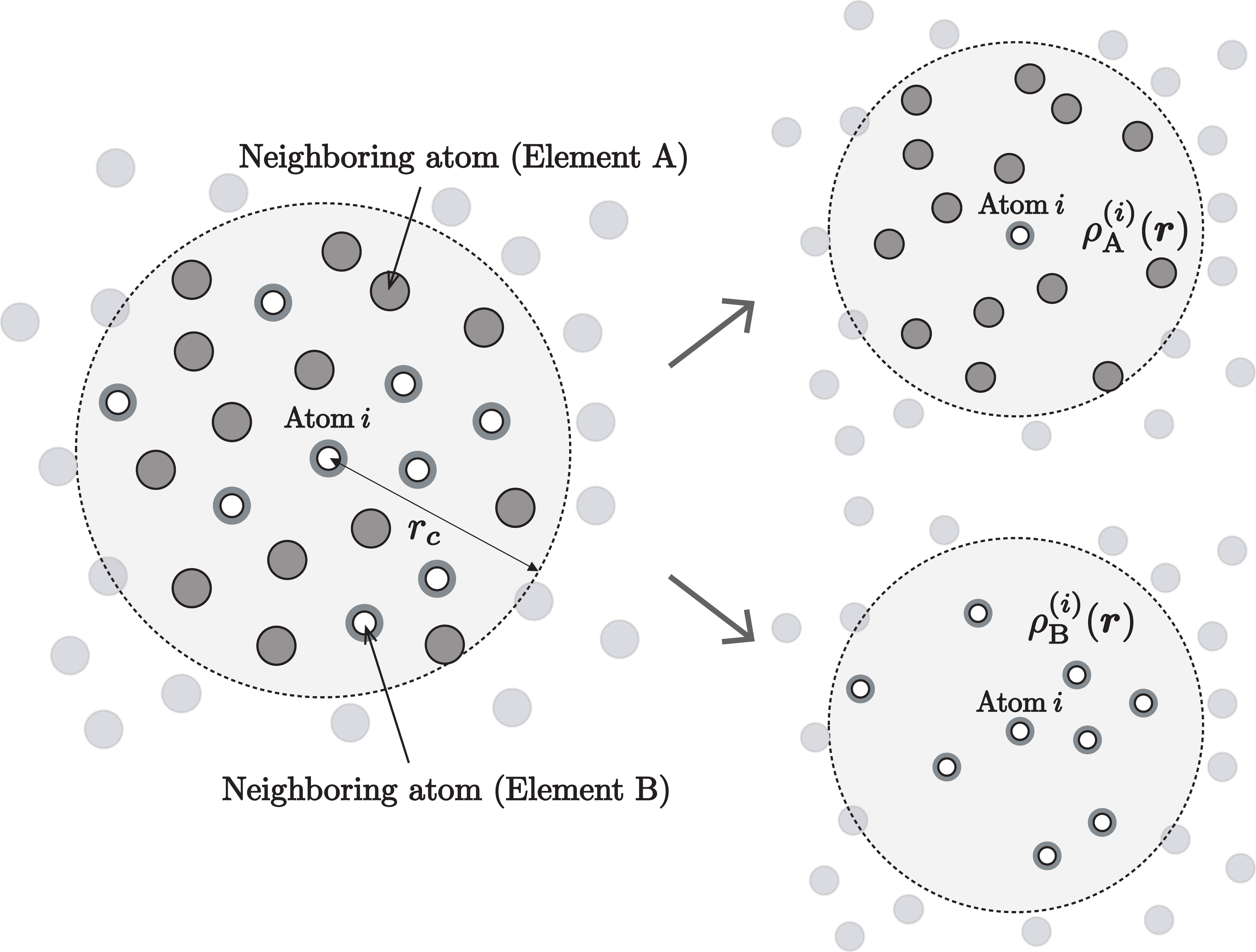}
\caption{
Schematic illustration of the neighboring atomic density around atom $i$ in a binary structure. 
Its decomposition into the neighboring atomic densities of elements A and B around atom $i$ is also shown.
}
\label{mlip-TiAl:atomic-distribution-schematic}
\end{figure}

\section{Potential energy models}
\label{mlip-TiAl:Sec-Model}

This section shows an extension of polynomial models for the potential energy proposed to develop MLPs for elemental systems \cite{PhysRevB.99.214108}.
This section is composed of a general description of the potential energy that is useful for deriving new potential energy models for a multicomponent system, a systematic set of structural features representing the neighboring atomic densities, and polynomial models for the potential energy used in this study.

\subsection{General description of potential energy}

Given a cutoff radius $r_c$, the short-range part of the potential energy for a structure, $E$, may be decomposed as
\begin{equation}
E = \sum_i E^{(i)},
\end{equation}
where $E^{(i)}$ denotes the contribution of atom $i$ within cutoff radius $r_c$.
The atomic contribution to the potential energy can be referred to as the atomic energy.
The atomic energy is then assumed to be expressed in a functional form of the neighboring atomic densities.
Figure \ref{mlip-TiAl:atomic-distribution-schematic} shows a schematic illustration of the neighboring atomic density around atom $i$ within cutoff radius $r_c$ and its decomposition into the neighboring atomic densities of the elements.
In a multicomponent system composed of elements $\{\rm A, \rm B, \cdots\}$, the atomic energy is written using functional $\mathcal{F}$ dependent on the element of atom $i$ as
\begin{equation}
E^{(i)} = \mathcal{F}_{s_i} \left[ \rho^{(i)}_{(s_i, \rm A)} , \rho^{(i)}_{(s_i, \rm B)}, \cdots \right],
\end{equation}
where $\rho^{(i)}_{(s_i,s)}$ denotes the neighboring atomic density of element $s$ ($s \in \{{\rm A},{\rm B},\cdots\}$) around atom $i$ of element $s_i$.

Subsequently, the neighboring atomic density of element $s$ around atom $i$ is expanded in terms of a basis set, because the expansion enables the functional form to be replaced with a function of its expansion coefficients.
For a given basis set $\{b_n\}$, the neighboring atomic densities can be expanded as
\begin{eqnarray}
\rho^{(i)}_{(s_i, {\rm A})}(\bm{r}) &=& \sum_n a_{n,(s_i,{\rm A})}^{(i)} b_n(\bm{r}) \nonumber \\
\rho^{(i)}_{(s_i, {\rm B})}(\bm{r}) &=& \sum_n a_{n,(s_i,{\rm B})}^{(i)} b_n(\bm{r}) \\
& \vdots & \nonumber
\end{eqnarray}
where $a_{n,(s_i,s)}^{(i)}$ denotes an order parameter characterizing the neighboring atomic density of element $s$ around atom $i$ of element $s_i$.
Using the order parameters, the atomic energy may be rewritten as
\begin{equation}
E^{(i)} = F'_{s_i} 
\left(
a_{1,(s_i,\rm A)}^{(i)}, a_{1,(s_i,\rm B)}^{(i)},\cdots,
a_{2,(s_i,\rm A)}^{(i)}, a_{2,(s_i,\rm B)}^{(i)},\cdots 
\right).
\label{mlip-TiAl:eqn-function1}
\end{equation}

Although function $F'$ of Eqn. (\ref{mlip-TiAl:eqn-function1}) depends on the element of atom $i$, it is convenient to introduce a unified function that is independent of the element by combining functions $F'$ for all elements.
In this study, a unified function for the atomic energy is formulated using order parameters defined for unordered pairs of elements.
This means that order parameter $a_{n,(s_i,s_j)}^{(i)}$ of atom $i$ and its swapped order parameter $a_{n,(s_j,s_i)}^{(j)}$ of atom $j$ are considered as the same variable in the unified function.
They are represented by $a_{n,\{s_i,s_j\}}^{(i)}$ and $a_{n,\{s_i,s_j\}}^{(j)}$, respectively.
Defining order parameter $a_{n,\{s_1,s_2\}}^{(i)}$ to be zero if $s_i$ is not included in $\{s_1,s_2\}$ ($s_i \notin \{s_1,s_2\}$), the atomic energy is written in an independent form of the element of atom $i$ as
\begin{eqnarray}
E^{(i)} &=& F'' \left(a_{1,\{\rm A,\rm A\}}^{(i)}, a_{1,\{\rm A,\rm B\}}^{(i)}, 
a_{1,\{\rm B,\rm B\}}^{(i)},\cdots, \right.\nonumber \\
&& \left. a_{2,\{\rm A,\rm A\}}^{(i)}, a_{2,\{\rm A,\rm B\}}^{(i)}, 
a_{2,\{\rm B,\rm B\}}^{(i)},\cdots\right),
\end{eqnarray}
where all combinations with the replacement of elements are enumerated for each $n$.

Moreover, an arbitrary rotation leaves the atomic energy invariant, although it generally changes the neighboring atomic densities and their order parameters.
Therefore, the atomic energy is required to be a function of O(3) invariants $\{d_{n'}^{(i)}\}$ as
\begin{equation}
E^{(i)} = F \left( d_1^{(i)}, d_2^{(i)}, \cdots \right),
\end{equation}
where the invariants are derived from the order parameters $\{a_{n,\{s_1,s_2\}}^{(i)}\}$.
Hereafter, the invariants representing the neighboring atomic density are referred to as ``structural features''.
The present formulation is useful for deriving potential energy models, the accuracy and computational efficiency of which can be controlled by the selections of structural features and function $F$.

\subsection{Structural features}
In this study, the neighboring atomic densities are expanded in terms of a basis set composed of radial functions or a basis set composed of products of radial functions and spherical harmonics \cite{bartok2013representing,PhysRevB.99.214108}, although it is also possible to use other basis sets in principle.
When the neighboring atomic density is expanded in terms of radial functions $\{f_n\}$, the neighboring atomic density around atom $i$ is expressed as
\begin{equation}
\rho^{(i)}_{(s_i,s)} (r) = \sum_{n} a^{(i)}_{n,\{s_i,s\}} f_n(r),
\end{equation}
where $r$ denotes the distance from the position of atom $i$.
Since order parameter $a^{(i)}_{n,\{s_i,s\}}$ is invariant for the O(3) group, it can be a pairwise structural feature denoted as
\begin{equation}
d^{(i)}_{n0,t} = a^{(i)}_{n,t},
\end{equation}
where $t$ identifies the unordered pair of elements, i.e., $t \in \{\{A,A\},\{A,B\},\cdots\}$.

When the neighboring atomic density is expanded in terms of products of radial functions $\{f_n\}$ and spherical harmonics $\{Y_{lm}\}$, the neighboring atomic density of element $s$ at a position $(r, \theta, \phi)$ in spherical coordinates centered at the position of atom $i$ is expressed as
\begin{equation}
\rho^{(i)}_{(s_i,s)} (r, \theta, \phi) = \sum_{nlm} a^{(i)}_{nlm,\{s_i,s\}} f_n(r) Y_{lm} (\theta, \phi),
\end{equation}
where order parameter $a^{(i)}_{nlm,\{s_i,s\}}$ is component $nlm$ of the neighboring atomic density.
Since the order parameters are not generally invariant for the O(3) group, polynomial invariants of the order parameters are adopted as structural features.
A $p$th-order polynomial invariant for a radial index $n$ and a set of pairs composed of the angular number and the element unordered pair $\{(l_1,t_1),(l_2,t_2),\cdots,(l_p,t_p)\}$ is defined as a linear combination of products of $p$ order parameters, expressed as
\begin{widetext}
\begin{equation}
\label{Eqn-invariant-form}
d_{nl_1l_2\cdots l_p,t_1t_2\cdots t_p,(\sigma)}^{(i)} = 
\sum_{m_1,m_2,\cdots, m_p} c^{l_1l_2\cdots l_p,(\sigma)}_{m_1m_2\cdots m_p} 
a_{nl_1m_1,t_1}^{(i)} a_{nl_2m_2,t_2}^{(i)} \cdots a_{nl_pm_p,t_p}^{(i)}.
\end{equation}
\end{widetext}

Nonzero polynomial invariants are derived from the sets satisfying the condition that the intersection of $t_p$ is not the empty set,
\begin{equation}
\prod_p t_p \neq \varnothing.
\end{equation}
Therefore, possible sets of pairs composed of the angular number and the element unordered pair $\{(l_1,t_1),(l_2,t_2),\cdots,(l_p,t_p)\}$ are enumerated for a given maximum angular number to obtain the entire set of polynomial invariants.
An example where the intersection of element pairs becomes the empty set is the case that a polynomial invariant is composed of order parameters with $t_1 = {\rm \{A,A\}}$ and $t_2 = {\rm \{B,B\}}$.
Such polynomial invariants are eliminated.

A coefficient set $\{c^{l_1l_2\cdots l_p,(\sigma)}_{m_1m_2\cdots m_p}\}$ is independent of the radial index $n$ and the element unordered pair $t$. 
Therefore, linearly independent coefficient sets are obtained using the group-theoretical projector operation method for a given set $\{l_1,l_2,\cdots,l_p\}$, as proposed in Ref. \onlinecite{PhysRevB.99.214108}, ensuring that the linear combinations are invariant for arbitrary rotation.
In terms of fourth- and higher-order polynomial invariants, multiple invariants are linearly independent for most of the set $\{l_1,l_2,\cdots,l_p\}$, which are distinguished by index $\sigma$ if necessary.
Note that the second- and third-order invariants are equivalent to a multicomponent extension of the angular Fourier series and the bispectrum reported in the literature, respectively \cite{kondor2008group,bartok2013representing}.

In this study, a finite set of Gaussian-type functions is adopted as radial functions in basis sets to expand the neighboring atomic density, expressed as
\begin{equation}
f_{n}(r)=\exp\left[-\beta_n(r-r_n)^{2}\right] f_c(r),
\end{equation}
where $\beta_n$ and $r_n$ are parameters.
The cutoff function $f_c$ is given by a cosine-based function as
\begin{eqnarray}
f_c(r) = \left\{
\begin{aligned}
& \frac{1}{2} \left[ \cos \left( \pi \frac{r}{r_c} \right) + 1\right] & (r \le r_c)\\
& 0 & (r > r_c)
\end{aligned}
\right ..
\end{eqnarray}
The order parameters of atom $i$ and element pair $\{s_i,s\}$ are approximately estimated from the neighboring atomic density of element $s$ around atom $i$ as
\begin{equation}
a_{nlm,\{s_i,s\}}^{(i)} = \sum_{\{j | r_{ij} \leq r_c,s_j = s\} }
f_n(r_{ij}) Y_{lm}^* (\theta_{ij}, \phi_{ij}),
\label{EquationOrderParameters}
\end{equation}
where $(r_{ij}, \theta_{ij}, \phi_{ij})$ denotes the spherical coordinates of neighboring atom $j$ centered at the position of atom $i$.
Although the Gaussian-type radial functions are not orthonormal, such an approximation of the order parameters is acceptable in the present polynomial-based framework, as also discussed in Ref. \onlinecite{PhysRevB.99.214108}.

\subsection{Polynomial models for atomic energy}
\label{mlip-TiAl:sec-polynomial-models}
Polynomial models are here employed to represent the atomic energy as a function of structural features.
Given a set of structural features $D = \set{d_1,d_2,\cdots}$, polynomial functions are written as
\begin{eqnarray}
f_1 \left(D\right) &=& \sum_{i'} w_{i'} d_{i'} \nonumber \\
f_2 \left(D\right) &=& \sum_{\{i',j'\}} w_{i'j'} d_{i'} d_{j'} \\
f_3 \left(D\right) &=& \sum_{\{i',j',k'\}} w_{i'j'k'} d_{i'} d_{j'} d_{k'} \nonumber \\
& \vdots & \nonumber
\end{eqnarray}
where $w$ denotes a regression coefficient.
Although the polynomial functions are described by all combinations of structural features, only nonzero polynomial terms are retained, which is analogous to the enumeration of nonzero polynomial invariants.
In a multicomponent system, a structural feature is composed of order parameters, each of which has an attribute on the element unordered pair $t$.
Therefore, when the element pair of the $p'$th order parameter in structure feature $d_{i'}$ of a polynomial term is denoted by $t_{i',p'}$, a nonzero polynomial term satisfies the condition that the intersection of $\{t_{i',p'}\}$ is not the empty set: 
\begin{equation}
\left( \prod_{p'} t_{i',p'} \right) \cap 
\left( \prod_{p'} t_{j',p'} \right) \cap \cdots \neq \varnothing.
\end{equation}
For example, the intersection of element pairs becomes the empty set if a polynomial term is composed of structural features with $t_{i',p_i'} = {\rm \{A,A\}}$ and $t_{j',p_j'} = {\rm \{B,B\}}$.
Such polynomial terms are eliminated from the polynomial functions.

The following polynomial models for the atomic energy are systematically applied to obtain Pareto optimal MLPs as described in Sec. \ref{mlip-TiAl:Sec-Pareto}.
The first model is a polynomial of pairwise structural features.
When a set of pairwise structural features is described as
\begin{equation}
D_{\rm pair}^{(i)} = \set{d_{n0,t}^{(i)}},
\end{equation}
the first model is expressed as
\begin{equation}
E^{(i)} = f_1\left(D_{\rm pair}^{(i)} \right) 
+ f_2\left(D_{\rm pair}^{(i)} \right) 
+ f_3\left(D_{\rm pair}^{(i)} \right) 
+ \cdots.
\end{equation}
The first model includes the special case that only powers of the pairwise structural features are considered, which was introduced for elemental systems in Refs. \onlinecite{PhysRevB.90.024101} and \onlinecite{PhysRevB.92.054113}.
The first model can also be regarded as a straightforward extension of embedded atom method (EAM) potentials \cite{PhysRevMaterials.1.063801}.

The second model for the atomic energy is a linear polynomial of polynomial invariants given by Eqn. (\ref{Eqn-invariant-form}). 
A set of polynomial invariants is described as
\begin{equation}
D^{(i)} = D_{\rm pair}^{(i)} \cup D_2^{(i)} 
\cup D_3^{(i)} \cup D_4^{(i)} \cup \cdots,
\end{equation}
where a set of $p$th-order polynomial invariants is denoted by
\begin{eqnarray}
D_2^{(i)} &=& \set{d_{nll,t_1t_2}^{(i)}} \nonumber \\
D_3^{(i)} &=& \set{d_{nl_1l_2l_3,t_1t_2t_3}^{(i)}} \\
D_4^{(i)} &=& \set{d_{nl_1l_2l_3l_4,t_1t_2t_3t_4,(\sigma)}^{(i)}} \nonumber.
\end{eqnarray}
A second-order invariant is identified with a single $l$ value because second-order linear combinations are invariant only when $l_1=l_2$ \cite{el-batanouny_wooten_2008,1987ltpt.book}.
The second model is then written as
\begin{equation}
E^{(i)} = f_1 \left( D^{(i)} \right),
\label{Eqn-linear-polynomial}
\end{equation}
which was introduced in Ref. \onlinecite{PhysRevB.99.214108} for elemental systems.
Note that a linear polynomial model with up to third-order invariants is equivalent to a spectral neighbor analysis potential (SNAP) \cite{Thompson2015316}, expressed as
\begin{equation}
E^{(i)} = f_1 \left( D_{\rm pair}^{(i)} \cup D_2^{(i)} \cup D_3^{(i)} \right).
\end{equation}

An extension of the second model is a polynomial of polynomial invariants described as
\begin{equation}
E^{(i)} = f_1 \left( D^{(i)} \right) + f_2 \left( D^{(i)} \right) 
+ f_3 \left( D^{(i)} \right) + \cdots.
\end{equation}
Note that a quadratic polynomial model of polynomial invariants up to the third order is equivalent to a quadratic SNAP \cite{doi:10.1063/1.5017641}.
Other extended models are also introduced, which are given by
\begin{eqnarray}
E^{(i)} &=& f_1 \left( D^{(i)} \right) + f_2 \left( D_{\rm pair}^{(i)} \right) 
+ f_3 \left( D_{\rm pair}^{(i)} \right) \nonumber \\
E^{(i)} &=& f_1 \left( D^{(i)} \right) 
+ f_2 \left( D_{\rm pair}^{(i)} \cup D_2^{(i)} \right) \\
E^{(i)} &=& f_1 \left( D^{(i)} \right) 
+ f_2 \left( D_{\rm pair}^{(i)} \cup D_2^{(i)} \cup D_3^{(i)} \right). \nonumber
\end{eqnarray}
They are decomposed into a linear polynomial of structural features and a polynomial of a subset of the structural features.

The atomic energy in all the models is measured from the sum of the energies of isolated atoms.
Moreover, the forces acting on atoms and the stress tensors in a multicomponent system can be derived in a similar manner to those in the elemental system derived in Ref. \onlinecite{PhysRevB.99.214108}.
Note that the above polynomial function forms were also applied to develop MLPs for elemental systems included in the \textsc{Machine Learning Potential Repository} \cite{MachineLearningPotentialRepositoryArxiv,MachineLearningPotentialRepository}.

\section{Datasets and computational procedure}
\label{mlip-TiAl:Sec-Datasets}

\begin{table*}[tbp]
\begin{ruledtabular}
\caption{
List of structure generators used for developing the training and test datasets.
}
\label{mlip-TiAl:table-structure-generators}
\begin{tabular}{cc|cc|cc}
ICSD CollCode & Structure type & ICSD CollCode & Structure type & ICSD CollCode & Structure type\\
\hline
239    & Cu$_3$Se$_2$   & 106786  & Hg$_2$Pt                   &  618295 & MoC$_{1-x}$             \\
5258   & FeSi$_2$    & 107998  & MoNi$_4$                   &  618702 & ScTe               \\
16504  & CrSi$_2$    & 108707  & HgMn                    &  625334 & Laves(2H)-MgZn$_2$                \\
16606  & Nb$_3$Te$_4$   & 108762  & Hg$_4$Pt                   &  626692 & NiAs             \\
30446  & Fe$_2$B     & 150584  & Fe$_{13}$Ge$_3$                 &  629380 & Al$_3$Os$_2$             \\
42428  & Fe$_3$Pt    & 155842  & Co$_5$Fe$_{11}$                 &  629406 & Cu$_4$Ti$_3$             \\
42472  & CoO      & 161109  & CoSn                    &  633467 & FeSe($tP2$)              \\
42773  & IrGe$_4$    & 161133  & Fe$_2$Si(HT)               &  635060 & FeSi               \\
52294  & GeTe     & 167735  & Ru$_2$B$_3$                   &  635208 & CoGa$_3$              \\
55492  & BaPt     & 168897  & LaI                     &  635642 & Hg$_5$Mn$_2$             \\
58471  & CuZr$_2$    & 169457  & ZrH$_2$                    &  638227 & CaF$_2$              \\
58607  & Au$_2$Ti    & 181127  & AuCu$_3$       &  639037 & HgIn               \\
58745  & Fe$_6$Ge$_6$Mg & 181788  & NaCl                    &  639148 & NiHg$_4$              \\
59508  & AuCu     & 185626  & Al$_3$Ni$_2$                  &  639227 & Si$_2$U$_3$              \\
59586  & Pd$_5$Th$_3$   & 188260  & Heusler-AlCu$_2$Mn         &  639879 & In$_5$In$_4$             \\
69199  & U$_3$Si     & 189695  & CuHg$_2$Ti                 &  640726 & CuSmP$_2$             \\
69557  & CdI$_2$($hP9$) & 189711  & Heusler-AlLiSi   &  643301 & Au$_3$Cd              \\
73839  & Ni$_3$S$_2$    &  240119 &  AlLi                   &  644708 & WC             \\
97006  & InMg$_2$    &  246555 &  Co$_2$Nd                  &  648572 & CuInPt$_2$                \\
99787  & Fe$_3$Pt    &  248490 &  Pt$_2$Si                  &  648748 & Pd$_4$Se              \\
100654 & BiSe     &  260285 &  UCl$_3$                   &  649037 & Ni$_3$Ti              \\
102712 & CoU      &  262070 &  AlLi($hP8$)              &  650527 & CsCl               \\
103775 & NaTl     &  409859 &  La$_2$Sb                  &  652553 & AlCr$_2$-MoSi$_2$                \\
103995 & Ga$_3$Ti$_2$   &  416747 &  Al$_3$Zr                  &  655706 & Cu$_2$Te(HT)              \\
104506 & Ni$_3$Sn    &  420250 &  LiPd$_2$Tl                &  659806 & GeTe              \\
105191 & Al$_3$Ti    &  424636 &  MnGa$_4$                  &  659829 & Al$_2$Li$_3$             \\
105521 & Al$_5$W     &  609153 &  AlPt$_3$                  &  659856 & LiPt               \\
105636 & PbU      &  610464 &  PbClF/Cu$_2$Sb               \\
105726 & Pd$_5$Ti$_3$   &  611176 &  Fe$_2$P              \\
105948 & InNi$_2$    &  611457 &  NbAs              \\
106325 & BiIn     &  611618 &  TiAs              \\
\end{tabular}
\end{ruledtabular}
\end{table*}

A straightforward development of training and test datasets for the Ti-Al binary system begins with a set of structure generators.
In this study, prototype structures reported as binary alloy entries in the Inorganic Crystal Structure Database (ICSD) \cite{bergerhoff1987crystal} are used as structure generators such that the datasets can cover a wide variety of structures.
Moreover, the prototype structures are restricted to those represented by unit cells with up to eight atoms.
A structure generator made by swapping elements in each prototype structure is also introduced; hence, the total number of structure generators is 150.
The structure generators are listed in Table \ref{mlip-TiAl:table-structure-generators}.

Given the structure generators, the atomic positions and lattice constants of the structure generators are fully optimized by DFT calculation to obtain their equilibrium structures.
A structure in the datasets is then generated by introducing random lattice expansion, random lattice distortion, and random atomic displacements into a supercell of each of the equilibrium structures. 
A mathematical description of the procedure can be found in Ref. \onlinecite{PhysRevB.99.214108}.
By repeatedly applying the procedure to the structure generators, 27,394 binary structures are generated for the datasets.
In addition to the binary structures and the equilibrium structures of the structure generators, existing data for elemental Ti and Al \cite{MachineLearningPotentialRepositoryArxiv} are also included in the present datasets.

For the total of 41,508 structures, DFT calculations are performed using the plane-wave-basis projector augmented wave (PAW) method \cite{PAW1} within the Perdew--Burke--Ernzerhof exchange-correlation functional \cite{GGA:PBE96} as implemented in the \textsc{vasp} code \cite{VASP1,VASP2,PAW2}.
The cutoff energy is set to 300 eV.
The total energies converge to less than 10$^{-3}$ meV/supercell.
The atomic positions and lattice constants for the structure generators are optimized until the residual forces are less than 10$^{-2}$ eV/\AA.

The regression coefficients of a potential energy model are estimated by linear ridge regression. 
The DFT total energies, the DFT forces acting on atoms, and the DFT stress tensors of structures in the training dataset are simultaneously used to estimate the regression coefficients, as adopted in Refs. \onlinecite{PhysRevB.99.214108} and \onlinecite{MachineLearningPotentialRepositoryArxiv}.
Therefore, the total number of training data reaches 5,178,510.

\section{Results and discussion}
\label{mlip-TiAl:Sec-Results}

\subsection{Pareto optimal MLPs}
\label{mlip-TiAl:Sec-Pareto}

\begin{figure}[tbp]
\includegraphics[clip,width=\linewidth]{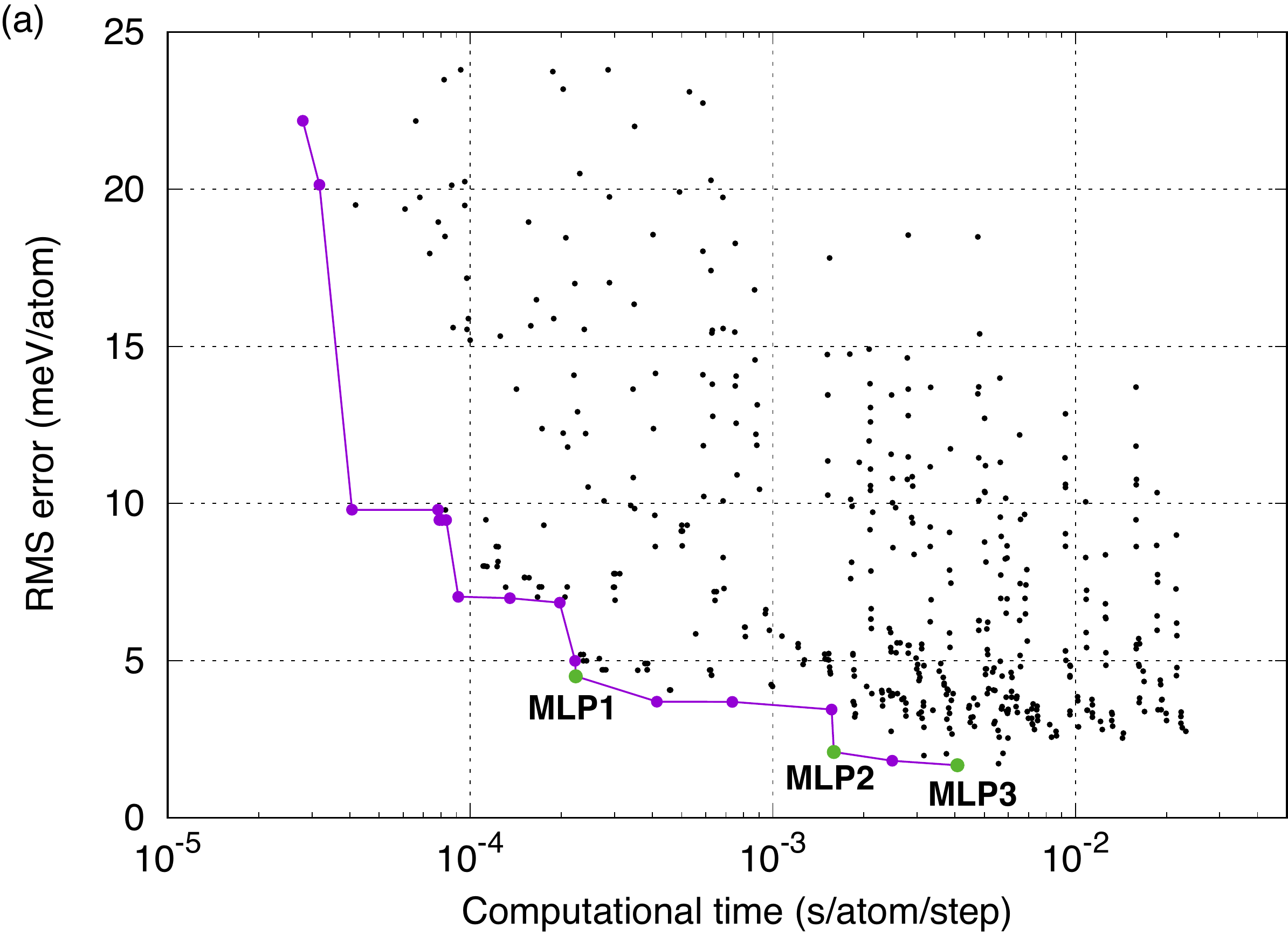}
\includegraphics[clip,width=\linewidth]{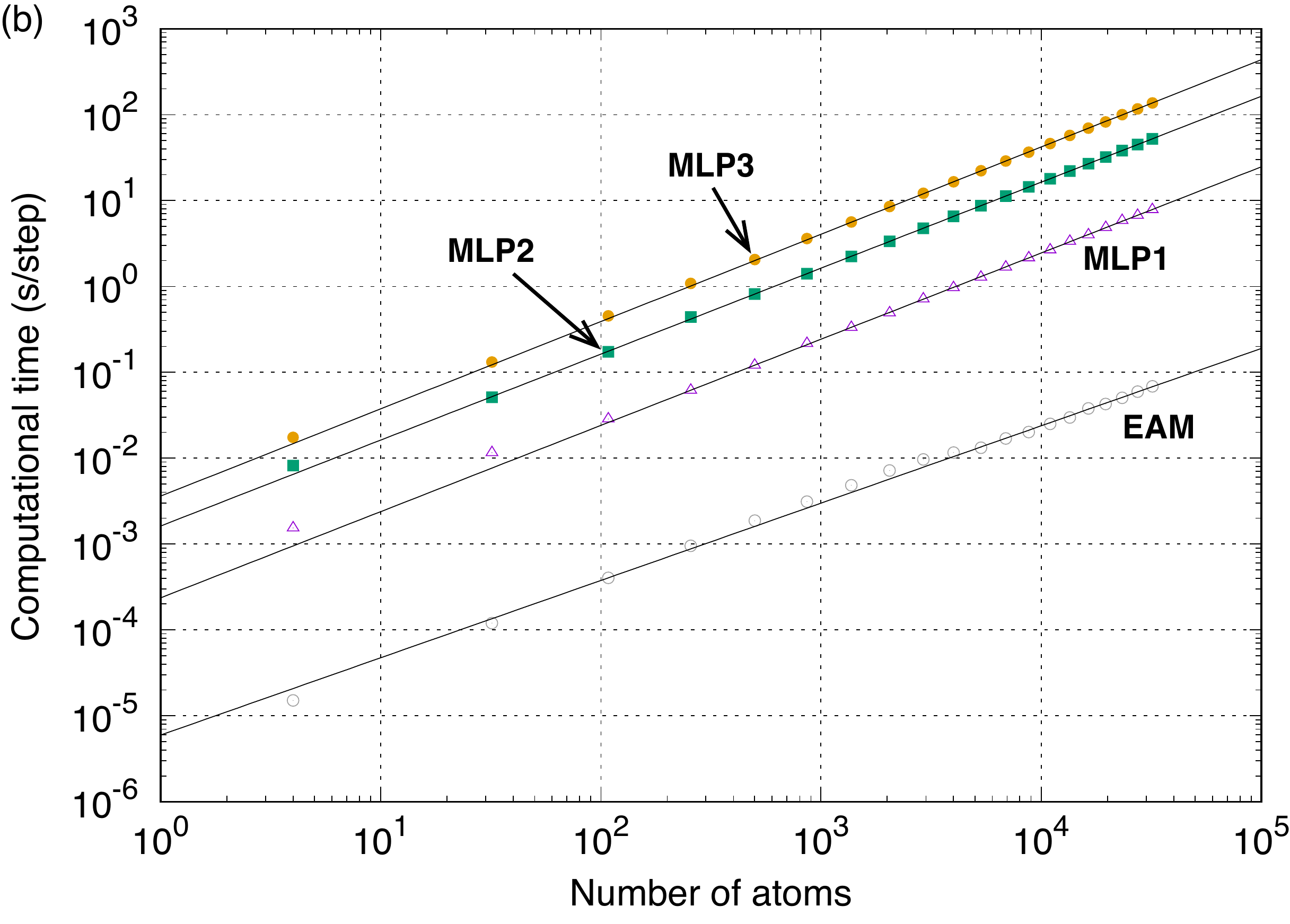}
\caption{
(a) Distribution of MLPs for the Ti-Al binary system.
The purple closed circles show Pareto optimal points of the distribution with different trade-offs between accuracy and computational efficiency. 
The green closed circles indicate the MLP with the lowest prediction error denoted by ``MLP3'' and the Pareto optimal MLPs with high computational cost performance denoted by ``MLP1'' and ``MLP2''.
The computational time indicates the elapsed time for a single-point calculation normalized by the number of atoms.
The elapsed time is measured using a single core of Intel\textregistered\ Xeon\textregistered\ E5-2695 v4 (2.10 GHz).
(b) Dependence of the computational time required for a single-point calculation on the number of atoms.
The dependence of the computational time using the EAM potential \cite{PhysRevB.68.024102} is also shown for comparison.
}
\label{mlip-TiAl:fig-Pareto}
\end{figure}

\begin{table}[tbp]
\begin{ruledtabular}
\caption{
Model parameters and prediction errors of MLP1, MLP2, and MLP3 for the Ti-Al binary system.
}
\label{mlip-TiAl:model-parameter}
\begin{tabular}{lccc}
& MLP1 & MLP2 & MLP3 \\
\hline
Number of coefficients & 7875 & 27,520 & 61,605 \\
RMS error (energy, meV/atom)  & 4.51  & 2.09  & 1.67  \\
RMS error (force, eV/\AA)  & 0.138  & 0.074  & 0.066  \\
Cutoff radius (\AA) & 8.0 & 6.0 & 8.0 \\
Number of radial functions & 10 & 10 & 15 \\
Polynomial order (function $F$) & 2 & 2 & 2 \\
Polynomial order (invariants) & 3 & 2 & 2 \\
$\set{l_{\rm max}^{(2)}, l_{\rm max}^{(3)},\cdots}$ & [0,0] & [4] & [4] \\
\end{tabular}
\end{ruledtabular}
\end{table}

The polynomial models given in Sec. \ref{mlip-TiAl:sec-polynomial-models} are systematically applied to develop MLPs in the Ti-Al binary system.
Because the accuracy and computational efficiency of the MLPs strongly depend on several input parameters, a systematic grid search is performed to find their optimal values.
The parameters in the grid search are
the cutoff radius,
the type of structural features,
the type of potential energy model,
the number of radial functions,
the polynomial order in the potential energy model,
and the truncation of the polynomial invariants, i.e., the maximum angular numbers of spherical harmonics, $\{l_{\rm max}^{(2)}, l_{\rm max}^{(3)}, \cdots, l_{\rm max}^{(p_{\rm max})}\}$, and the polynomial order of the invariants, $p_{\rm max}$.

Figure \ref{mlip-TiAl:fig-Pareto} (a) shows the distribution of MLPs obtained from the grid search.
The root mean squared (RMS) error for the test dataset is used as an estimator of the accuracy of MLPs.
The computational time indicates the elapsed time for a single point calculation normalized by the number of atoms.
Figure \ref{mlip-TiAl:fig-Pareto} (a) also shows the Pareto optimal MLPs when optimizing both the accuracy and computational efficiency simultaneously.
As can be seen in Fig. \ref{mlip-TiAl:fig-Pareto} (a), the accuracy and computational efficiency of MLPs are conflicting properties; hence, the Pareto optimal MLPs can be optimal ones with different trade-offs.

In performing an atomistic simulation, an appropriate MLP must be chosen from the Pareto optimal ones according to its target system and purpose.
Therefore, a convenient score that can estimate the computational cost performance is required to find an MLP with high computational cost performance in a simplified manner.
In this study, functions $t + \Delta E$ and $10t + \Delta E$ that should be minimized are introduced, where $t$ and $\Delta E$ denote the computational time with the unit of ms/atom/step and the RMS error with the unit of meV/atom, respectively.

Figure \ref{mlip-TiAl:fig-Pareto} (a) shows the MLP with the lowest RMS error and two Pareto optimal MLPs showing high computational cost performance.
The MLP with the lowest RMS error is denoted by ``MLP3'', whereas the two MLPs showing high computational cost performance are denoted by ``MLP1'' and ``MLP2''.
MLP1 and MLP2 are obtained by minimizing the scores $10t + \Delta E$ and $t + \Delta E$, respectively. 
As can be seen in Fig. \ref{mlip-TiAl:fig-Pareto} (a), they exhibit high computational efficiency without significantly increasing the RMS error.
The model parameters and RMS errors of MLP1, MLP2, and MLP3 are listed in Table \ref{mlip-TiAl:model-parameter}.
The RMS errors of MLP2 and MLP3 are close, 2.09 and 1.67 meV/atom, respectively.
The RMS error of MLP1 is 4.51 meV/atom, which is greater than those of MLP2 and MLP3, whereas MLP1 is ten times more computationally efficient than MLP2.
As described in Table \ref{mlip-TiAl:model-parameter}, all the MLPs are derived from polynomial invariants. 
However, MLP1 is developed only using the $l=0$ component of spherical harmonics.
Therefore, MLP1 can be regarded as a polynomial model of pairwise structural features.

Figure \ref{mlip-TiAl:fig-Pareto} (b) shows the computational time required for a single point calculation using the EAM potential \cite{PhysRevB.68.024102}, MLP1, MLP2, and MLP3.
The computational time is evaluated for structures with up to 32,000 atoms constructed by the expansions of the four-atom unit cell of a AuCu-type ($L1_0$) structure with a lattice constant of 4 \AA.
The computational time is measured by implementing the present MLPs \cite{LammpsMLIPpackage} in the \textsc{lammps} code \cite{lammps,plimpton1995fast}.
As can be seen in Fig. \ref{mlip-TiAl:fig-Pareto} (b), the EAM potential and the MLPs show linear scaling of the computational time with respect to the number of atoms.
Therefore, the computational time normalized by the number of atoms can be an estimator of the computational efficiency of MLPs, and the computational time required for a simulation of $n_{\rm step}$ steps using a structure with $n_{\rm atom}$ atoms can be estimated as $ t \times n_{\rm atom} \times n_{\rm step}$.

\subsection{Cohesive energy}

\begin{figure}[tbp]
\includegraphics[clip,width=\linewidth]{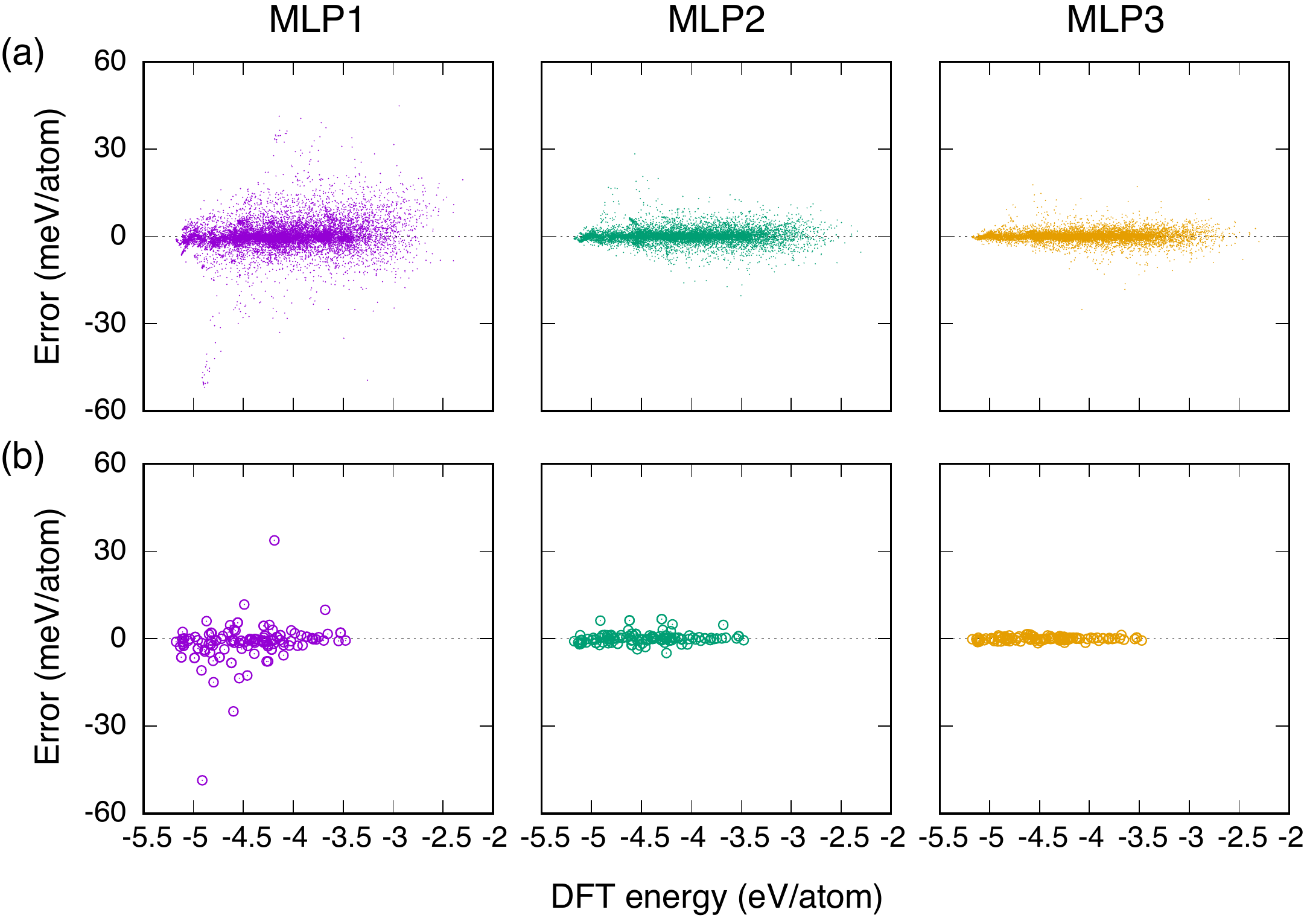}
\caption{
(a) Distribution of the prediction errors for structures in the training and test datasets.
(b) Distribution of the prediction errors for the equilibrium structure generators listed in Table \ref{mlip-TiAl:table-structure-generators}.
}
\label{mlip-TiAl:error-distribution}
\end{figure}

Figure \ref{mlip-TiAl:error-distribution} (a) shows the distribution of the prediction errors for structures in the training and test datasets. 
The degree of scattering in the distribution decreases in the order of MLP1, MLP2, and MLP3, which coincides with the decreasing order of the RMS errors shown in Table \ref{mlip-TiAl:model-parameter}.
Figure \ref{mlip-TiAl:error-distribution} (b) shows the distribution of the prediction errors for the equilibrium structure generators listed in Table \ref{mlip-TiAl:table-structure-generators}.
The distribution of the prediction errors for the structure generators indicates the predictive power for the cohesive energy.
The prediction errors for some of the structure generators are significant in MLP1.
On the other hand, the prediction errors are trivial for all the structure generators in MLP2 and MLP3, which indicates that MLP2 and MLP3 should have high predictive power for the cohesive energy in a wide variety of binary ordered structures.

\subsection{Formation energy}

\begin{table}[tbp]
\begin{ruledtabular}
\caption{
Formation energies of selected ordered structures. (unit: meV/atom)
}
\label{mlip-TiAl:formation_energy}
\begin{tabular}{lccccc}
Structure type & EAM\footnote{Ref. \onlinecite{PhysRevB.68.024102}} & MLP1 & MLP2 & MLP3 & DFT \\
\hline
Ti$_5$Al \\
Al$_5$W                 & $-$180 & $-$156 & $-$135 & $-$134 & $-$136 \\
\hline
Ti$_4$Al \\
MoNi$_4$ ($D1_a$)       & $-$221 & $-$202 & $-$180 & $-$178 & $-$179 \\
\hline
Ti$_3$Al \\
Ni$_3$Sn ($D0_{19}$)    & $-$288 & $-$294 & $-$278 & $-$277 & $-$280 \\
Ni$_3$Ti ($D0_{24}$)    & $-$288 & $-$290 & $-$267 & $-$265 & $-$267 \\
AuCu$_3$ ($L1_2$)       & $-$288 & $-$283 & $-$263 & $-$262 & $-$264 \\
Al$_3$Zr ($D0_{23}$)    & $-$282 & $-$276 & $-$258 & $-$257 & $-$259 \\
Al$_3$Ti ($D0_{22}$)    & $-$275 & $-$275 & $-$253 & $-$252 & $-$254 \\
AlCu$_2$Mn ($L2_1$)     & $-$229 & $-$164 & $-$143 & $-$142 & $-$143 \\
\hline
Ti$_5$Al$_2$\\
Hg$_5$Mn$_2$            & $-$161 & $-$70  & $-$53  & $-$51  & $-$53 \\
\hline
Ti$_{11}$Al$_5$\\
Co$_5$Fe$_{11}$         & $-$246 & $-$254 & $-$235 & $-$234 & $-$236 \\
\hline
Ti$_2$Al \\
InNi$_2$ ($B8_2$)       & $-$208 & $-$317 & $-$305 & $-$304 & $-$305 \\
CuZr$_2$                & $-$299 & $-$284 & $-$268 & $-$267 & $-$269 \\
Fe$_2$P  ($C22$)        & $-$276 & $-$205 & $-$237 & $-$228 & $-$228 \\
CrSi$_2$ ($C40$)        & $-$245 & $-$210 & $-$187 & $-$185 & $-$186 \\
Fe$_2$B                 & $-$137 & $-$160 & $-$143 & $-$141 & $-$143 \\
Cu$_2$Sb ($C38$)        & $-$128 & $-$160 & $-$139 & $-$137 & $-$139 \\
FeSi$_2$                & $-$118 & $-$22  & $-$7   & $-$4   & $-$7 \\
\hline
Ti$_5$Al$_3$\\
Pd$_5$Th$_3$            & $-$255 & $-$204 & $-$192 & $-$191 & $-$193 \\
\hline
Ti$_3$Al$_2$ \\
Ga$_3$Ti$_2$            & $-$363 & $-$344 & $-$332 & $-$331 & $-$331 \\
Al$_3$Os$_2$            & $-$290 & $-$302 & $-$288 & $-$287 & $-$288 \\
Si$_2$U$_3$ ($D5_a$)    & $-$140 & $-$253 & $-$246 & $-$244 & $-$245 \\
\hline
Ti$_4$Al$_3$\\
Nb$_3$Te$_4$            & $-$340 & $-$340 & $-$327 & $-$327 & $-$329 \\
Cu$_4$Ti$_3$            & $-$281 & $-$232 & $-$217 & $-$216 & $-$217 \\
\hline
TiAl \\
AuCu ($L1_0$)           & $-$404 & $-$417 & $-$404 & $-$403 & $-$404 \\
PbU                     & $-$370 & $-$375 & $-$366 & $-$367 & $-$368 \\
CoU ($B_a$)             & $-$340 & $-$298 & $-$283 & $-$282 & $-$283 \\
CsCl ($B2$)             & $-$286 & $-$280 & $-$265 & $-$263 & $-$262 \\
NiAs ($B8_1$)           & $-$311 & $-$263 & $-$257 & $-$251 & $-$251 \\
WC ($B_h$)              & $-$296 & $-$255 & $-$250 & $-$249 & $-$250 \\
ScTe                    & $-$307 & $-$238 & $-$223 & $-$224 & $-$225 \\
BiSe                    & $-$265 & $-$185 & $-$172 & $-$172 & $-$173 \\
FeSi ($B20$)            & $-$206 & $-$145 & $-$118 & $-$118 & $-$120 \\
NaTl ($B32$)            & $-$319 & $-$83  & $-$70  & $-$69  & $-$71  \\
\hline
Ti$_2$Al$_3$ \\
Ga$_3$Ti$_2$            & $-$370 & $-$408 & $-$416 & $-$418 & $-$419 \\
Al$_3$Os$_2$            & $-$353 & $-$319 & $-$309 & $-$309 & $-$310 \\
Si$_2$U$_3$ ($D5_a$)    & $-$91  & $-$80  & $-$73  & $-$72  & $-$72 \\
\end{tabular}
\end{ruledtabular}
\end{table}

\begin{table}[tbp]
\begin{ruledtabular}
\caption{
Formation energies of selected ordered structures (continued). (unit: meV/atom)
}
\label{mlip-TiAl:formation_energy2}
\begin{tabular}{lccccc}
Structure type & EAM\footnote{Ref. \onlinecite{PhysRevB.68.024102}} & MLP1 & MLP2 & MLP3 & DFT \\
\hline
Ti$_3$Al$_5$ \\
Pd$_5$Th$_3$            & $-$239 & $-$332 & $-$322 & $-$324 & $-$325 \\
Pd$_5$Ti$_3$            & $-$340 & $-$293 & $-$285 & $-$284 & $-$284 \\
\hline
TiAl$_2$ \\
MgZn$_2$ ($C14$)        & $-$331 & $-$334 & $-$326 & $-$325 & $-$326 \\
Co$_2$Nd                & $-$313 & $-$323 & $-$315 & $-$314 & $-$315 \\
CuZr$_2$                & $-$320 & $-$252 & $-$244 & $-$243 & $-$245 \\
La$_2$Sb                & $-$181 & $-$248 & $-$238 & $-$238 & $-$239 \\
Cu$_2$Sb ($C38$)        & $-$184 & $-$239 & $-$232 & $-$229 & $-$230 \\
FeSi$_2$                & $-$156 & $-$212 & $-$210 & $-$210 & $-$212 \\
Fe$_2$P ($C22$)         & $-$232 & $-$206 & $-$191 & $-$192 & $-$195 \\
Hg$_2$Pt                & 183  & $-$45  & $-$40  & $-$41  & $-$43  \\
\hline
Ti$_5$Al$_{11}$ \\
Co$_5$Fe$_{11}$         & $-$331 & $-$380 & $-$373 & $-$373 & $-$374 \\
\hline
Ti$_2$Al$_5$ \\
Hg$_5$Mn$_2$            & $-$168 & $-$295 & $-$289 & $-$289 & $-$289 \\
\hline
TiAl$_3$ \\
Al$_3$Zr ($D0_{23}$)    & $-$297 & $-$407 & $-$402 & $-$402 & $-$403 \\
Al$_3$Ti ($D0_{22}$)    & $-$289 & $-$407 & $-$403 & $-$397 & $-$397 \\
AuCu$_3$ ($L1_2$)       & $-$303 & $-$375 & $-$369 & $-$369 & $-$370 \\
Ni$_3$Ti ($D0_{24}$)    & $-$293 & $-$349 & $-$338 & $-$338 & $-$338 \\
Ni$_3$Sn ($D0_{19}$)    & $-$286 & $-$321 & $-$318 & $-$318 & $-$319 \\
\hline
TiAl$_4$ \\
MoNi$_4$ ($D1_a$)       & $-$240 & $-$223 & $-$216 & $-$215 & $-$216 \\
\hline
TiAl$_5$ \\
Al$_5$W                 & $-$188 & $-$184 & $-$180 & $-$180 & $-$180 \\
\end{tabular}
\end{ruledtabular}
\end{table}

The formation energy of a given ordered structure is more challenging to predict accurately than its cohesive energy because high predictive power is required not only for the ordered structure but also for the reference structures of hexagonal-close-packed (hcp) Ti and face-centered-cubic (fcc) Al.
Furthermore, the local geometry relaxation for the ordered structure is crucial for evaluating the formation energy.
Therefore, MLPs are needed to derive an accurate potential energy surface around the initial and equilibrium structures.

Tables \ref{mlip-TiAl:formation_energy} and \ref{mlip-TiAl:formation_energy2} show the formation energies of selected ordered structures predicted using the EAM potential, MLP1, MLP2, and MLP3, compared with those predicted by DFT calculation.
Note that the ordered structures cover a wide range of compositions.
The RMS errors of the EAM potential, MLP1, MLP2, and MLP3 for the formation energy are 68.0, 13.8, 2.0, and 1.5 meV/atom, respectively.
This reveals that MLP2 and MLP3 have high predictive power for the formation energy in a wide range of structures.
The RMS error of MLP1 for the formation energy is significant as a consequence of the systematic deviation of the formation energy for the overall ordered structures.
As can be seen in Tables \ref{mlip-TiAl:formation_energy} and \ref{mlip-TiAl:formation_energy2}, the formation energies of most of the ordered structures predicted using MLP1 are approximately 10 meV/atom lower than those predicted by DFT calculation, which originates from the fact that the prediction error of MLP1 for hcp-Ti is significant ($+$27.4 meV/atom).
Moreover, the prediction error of the EAM potential is much more significant, and the EAM potential fails to reconstruct the hierarchy of the formation energies predicted by DFT calculation.

\subsection{Elastic constants}

\begin{table}[tbp]
\begin{ruledtabular}
\caption{
Lattice constants and elastic constants of TiAl ($L1_0$, $B2$, and $B8_1$).
}
\label{mlip-TiAl:elastic-constants-0.5}
\begin{tabular}{lccccc}
 & EAM\footnote{Ref. \onlinecite{PhysRevB.68.024102}} & MLP1 & MLP2 & MLP3 & DFT \\
\hline
$\gamma$-TiAl (CuAu, $L1_0$) \\
$a_0$ (\AA) & 2.827 & 2.812 & 2.812 & 2.812 & 2.813 \\
$c_0$ (\AA) & 4.187 & 4.080 & 4.078 & 4.080 & 4.079 \\
$C_{11}$ (GPa) & 237 & 219 & 195 & 190 & 195 \\
$C_{12}$ (GPa) & 67  & 35 & 63 & 65 & 66 \\
$C_{13}$ (GPa) & 114 & 87 & 90 & 90 & 89 \\
$C_{33}$ (GPa) & 213 & 189 & 176 & 176 & 173 \\
$C_{44}$ (GPa) & 92 & 112 & 114 & 114 & 113 \\
$C_{66}$ (GPa) & 45 & 52 & 39 & 39 & 38 \\
TiAl (CsCl, $B2$) \\
$a_0$ (\AA)    & 3.278  & 3.184 & 3.183 & 3.182 & 3.182 \\
$C_{11}$ (GPa) & 80     & 67    & 82    & 69    & 74    \\
$C_{12}$ (GPa) & 121    & 132   & 134   & 141   & 136   \\
$C_{44}$ (GPa) & 95     & 80    & 87    & 82    & 66   \\
TiAl (NiAs, $B8_1$) \\
$a_0$ (\AA) & 2.853 & 2.880 & 2.878 & 2.877 & 2.879 \\
$c_0$ (\AA) & 9.370 & 9.269 & 9.264 & 9.276 & 9.263 \\
$C_{11}$ (GPa) & 157 & 149 & 126 & 132 & 136 \\
$C_{12}$ (GPa) & 94  & 113 & 99  & 94  & 96  \\
$C_{13}$ (GPa) & 93  & 88  & 65  & 72  & 74  \\
$C_{33}$ (GPa) & 292 & 277 & 250 & 220 & 223 \\
$C_{44}$ (GPa) & 67  & 73  & 73  & 71  & 75  \\
$C_{66}$ (GPa) & 32  & 18  & 14  & 19  & 20  \\
\end{tabular}
\end{ruledtabular}
\end{table}

\begin{table}[htbp]
\begin{ruledtabular}
\caption{
Lattice constants and elastic constants of Ti$_3$Al ($D0_{19}$, $D0_{22}$, and $L1_2$).
}
\label{mlip-TiAl:elastic-constants-0.25}
\begin{tabular}{lccccc}
 & EAM\footnote{Ref. \onlinecite{PhysRevB.68.024102}} & MLP1 & MLP2 & MLP3 & DFT \\
\hline
Ti$_3$Al (Ni$_3$Sn, $D0_{19}$) \\
$a_0$ (\AA) & 5.784 & 5.728 & 5.731 & 5.729 & 5.726 \\
$c_0$ (\AA) & 4.750 & 4.646 & 4.643 & 4.644 & 4.646 \\
$C_{11}$ (GPa) & 199 & 187 & 181 & 196 & 195 \\
$C_{12}$ (GPa) & 89  & 113 & 104 & 99  & 90  \\
$C_{13}$ (GPa) & 74  & 61  & 67  & 71  & 70  \\
$C_{33}$ (GPa) & 224 & 237 & 235 & 231 & 232 \\
$C_{44}$ (GPa) & 51  & 44  & 47  & 63  & 59  \\
$C_{66}$ (GPa) & 55  & 37  & 39  & 48  & 53  \\
Ti$_3$Al (TiAl$_3$, $D0_{22}$) \\
$a_0$ (\AA) & 4.082 & 3.943 & 3.961 & 3.962 & 3.960 \\
$c_0$ (\AA) & 8.252 & 8.479 & 8.423 & 8.433 & 8.444 \\
$C_{11}$ (GPa) & 161 & 192 & 173 & 177 & 173 \\
$C_{12}$ (GPa) & 100 & 145 & 105 & 92  & 97  \\
$C_{13}$ (GPa) & 93  & 119 & 101 & 89  & 87  \\
$C_{33}$ (GPa) & 146 & 260 & 168 & 151 & 171 \\
$C_{44}$ (GPa) & 71  & 73  & 84  & 79  & 87  \\
$C_{66}$ (GPa) & 78  & 98  & 94  & 98  & 89  \\
Ti$_3$Al (Cu$_3$Au, $L1_2$) \\
$a_0$ (\AA) & 4.089 & 4.036 & 4.036 & 4.036 & 4.035 \\
$C_{11}$ (GPa) & 165 & 143 & 165 & 172 & 175 \\
$C_{12}$ (GPa) & 97  & 102 & 91  & 89  & 89  \\
$C_{44}$ (GPa) & 76  & 92  & 93  & 92  & 90  \\
\end{tabular}
\end{ruledtabular}
\end{table}

\begin{table}[htbp]
\begin{ruledtabular}
\caption{
Lattice constants and elastic constants of TiAl$_3$ ($D0_{22}$, $D0_{19}$, and $L1_2$).
}
\label{mlip-TiAl:elastic-constants-0.75}
\begin{tabular}{lccccc}
 & EAM\footnote{Ref. \onlinecite{PhysRevB.68.024102}} & MLP1 & MLP2 & MLP3 & DFT \\
\hline
TiAl$_3$ (TiAl$_3$, $D0_{22}$) \\
$a_0$ (\AA) & 4.049 & 3.838 & 3.842 & 3.842 & 3.841 \\
$c_0$ (\AA) & 8.139 & 8.626 & 8.621 & 8.609 & 8.608 \\
$C_{11}$ (GPa) & 170 & 190 & 213 & 188 & 196 \\
$C_{12}$ (GPa) & 98  & 100 & 76  & 81  & 87  \\
$C_{13}$ (GPa) & 89  & 2   & 26  & 38  & 47  \\
$C_{33}$ (GPa) & 140 & 212 & 192 & 223 & 220 \\
$C_{44}$ (GPa) & 62  & 70  & 83  & 83  & 95  \\
$C_{66}$ (GPa) & 71  & 140 & 126 & 120 & 129 \\
TiAl$_3$ (Ni$_3$Sn, $D0_{19}$) \\
$a_0$ (\AA) & 5.704 & 5.565 & 5.565 & 5.564 & 5.563 \\
$c_0$ (\AA) & 4.810 & 4.722 & 4.724 & 4.725 & 4.726 \\
$C_{11}$ (GPa) & 205 & 219 & 203 & 206 & 209 \\
$C_{12}$ (GPa) & 88  & 58  & 67  & 66  & 67  \\
$C_{13}$ (GPa) & 62  & 53  & 63  & 59  & 60  \\
$C_{33}$ (GPa) & 189 & 163 & 161 & 168 & 167 \\
$C_{44}$ (GPa) & 34  & 46  & 53  & 62  & 65  \\
$C_{66}$ (GPa) & 58  & 80  & 68  & 70  & 71  \\
TiAl$_3$ (Cu$_3$Au, $L1_2$) \\
$a_0$ (\AA) & 4.050 & 3.976 & 3.977 & 3.977 & 3.979 \\
$C_{11}$ (GPa) & 179 & 183 & 181 & 190 & 191 \\
$C_{12}$ (GPa) & 95  & 56  & 68  & 69  & 66  \\
$C_{44}$ (GPa) & 73  & 84  & 77  & 76  & 77  \\
\end{tabular}
\end{ruledtabular}
\end{table}

Tables \ref{mlip-TiAl:elastic-constants-0.5}, \ref{mlip-TiAl:elastic-constants-0.25}, and \ref{mlip-TiAl:elastic-constants-0.75} show the lattice constants and the elastic constants of nine structures, i.e., TiAl ($L1_0$), TiAl ($B2$), TiAl ($B8_1$), Ti$_3$Al ($D0_{19}$), Ti$_3$Al ($D0_{22}$), Ti$_3$Al ($L1_2$), TiAl$_3$ ($D0_{22}$), TiAl$_3$ ($D0_{19}$), and TiAl$_3$ ($L1_2$), which are predicted using the EAM potential, MLP1, MLP2, and MLP3, along with the lattice constants and the elastic constants obtained by DFT calculation.
The lattice constants predicted using the EAM potential deviate from those obtained by DFT calculation in most of the structures.
This deviation may arise from the fact that the EAM potential was developed by fitting to experimental lattice constants \cite{PhysRevB.68.024102}, excluding the descriptive power of the EAM potential model.
In contrast, the MLPs can compute the lattice constants accurately in all of the structures.
Regarding the elastic constants, the elastic constants predicted using the EAM potential and MLP1 are close to those obtained by DFT calculation in many structures; however, the EAM potential and MLP1 fail to predict the elastic constants accurately in a few structures.
Also, the elastic constants predicted using MLP2 and MLP3 are almost the same as those obtained by DFT calculation in all structures.

\subsection{Phonon properties}

\begin{figure*}[tbp]
\includegraphics[clip,width=0.85\linewidth]{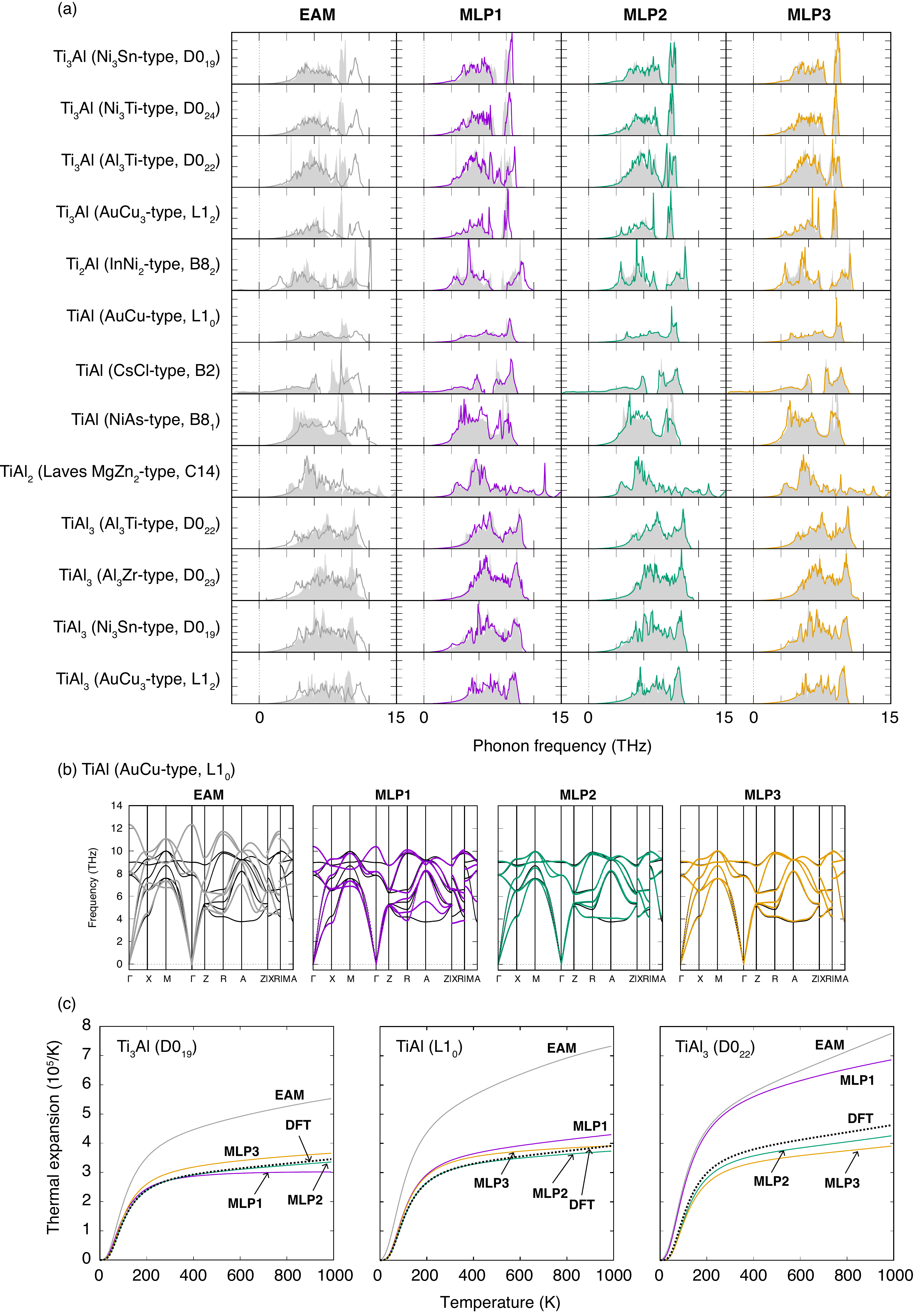}
\caption{
(a) Phonon density of states for 13 selected compounds in the Ti-Al binary system, predicted using the EAM potential and the MLPs.
The shaded region indicates the phonon density of states computed by DFT calculation.
(b) Phonon dispersion curves for $\gamma$-TiAl predicted using the EAM potential and the MLPs.
The solid black lines indicate the phonon dispersion curves predicted by DFT calculation.
(c) Temperature dependence of the thermal expansion predicted using the EAM potential and the MLPs.
The broken black lines show the thermal expansion computed by DFT calculation.
}
\label{mlip-TiAl:Fig-phonon}
\end{figure*}

The predictive power of the MLPs for phonon properties and thermal expansion is examined.
The phonon properties and thermal expansion are calculated using a finite displacement method implemented in the \textsc{phonopy} code \cite{phonopy2}.
Figure \ref{mlip-TiAl:Fig-phonon} (a) shows the phonon density of states for 13 structures predicted using the EAM potential, MLP1, MLP2, and MLP3.
The supercells required to compute the phonon properties are constructed by the expansions of the conventional unit cells of the structures.
The number of atoms included in the supercells ranges from 54 to 162.
The EAM potential predicts the phonon density of states well in the low-frequency region for many structures, while the deviation from the DFT phonon density of states is large in the high-frequency region.
Conversely, the phonon density of states predicted using the MLPs and those predicted by DFT calculation overlap for all the structures, particularly those predicted using MLP2 and MLP3.

Figure \ref{mlip-TiAl:Fig-phonon} (b) shows the phonon dispersion curves for TiAl (AuCu-type, $L1_0$) predicted using the EAM potential and the MLPs, compared with those predicted by DFT calculation.
The deviation of the EAM phonon dispersions from the DFT phonon dispersions is significant in the high-frequency region.
On the other hand, the phonon dispersions of MLP2 and MLP3 are consistent with those of DFT calculation.

Figure \ref{mlip-TiAl:Fig-phonon} (c) shows the temperature dependence of the thermal expansion, calculated using a quasi-harmonic approximation, in Ti$_3$Al ($D0_{19}$), TiAl ($L1_0$), and TiAl$_3$ ($D0_{22}$), which are experimentally observed in the Ti-Al binary system.
As can be seen in Fig. \ref{mlip-TiAl:Fig-phonon} (c), the thermal expansion of the EAM potential differs from that of the DFT calculation in all the structures.
On the other hand, MLP2 and MLP3 derive the temperature dependence of the thermal expansion accurately in all the structures, even though the thermal expansion is more challenging to predict accurately than the phonon density of states and the phonon dispersion curves.
The accurate prediction of the thermal expansion indicates that MLP2 and MLP3 can accurately evaluate the volume dependence of the whole range of phonon frequencies.

\subsection{Bain path between $\bm{\gamma}$-TiAl and $\bm{B2}$}

\begin{figure}[tbp]
\includegraphics[clip,width=\linewidth]{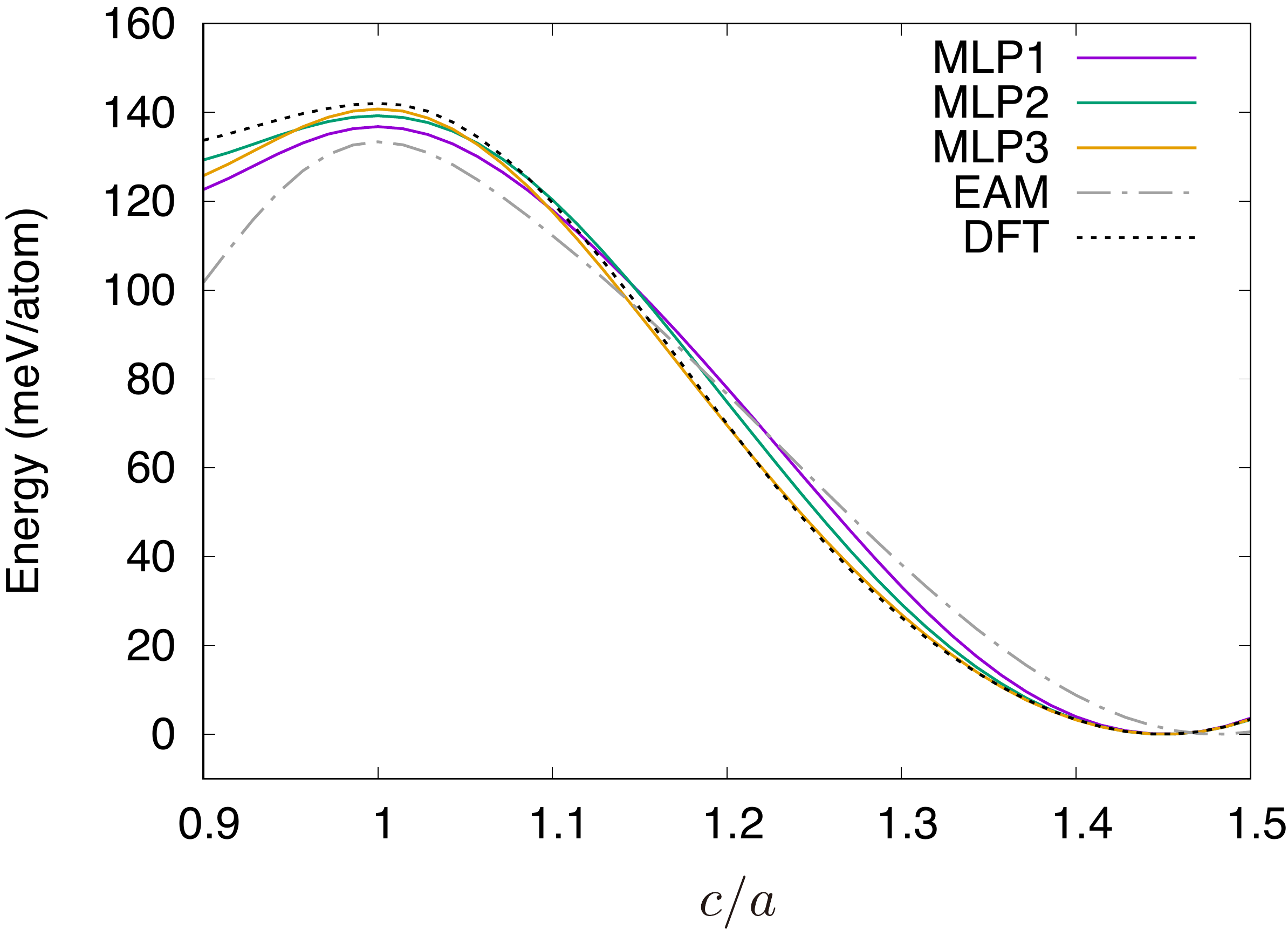}
\caption{
Energy profiles along the Bain path between the $L1_0$ structure and the $B2$ structure.
}
\label{mlip-TiAl:Fig-bain-path}
\end{figure}

The energy profile along the Bain path between the $\gamma$-TiAl ($L1_0$) structure and the $B2$ structure is calculated using the EAM potential and the MLPs, and compared with that obtained by DFT calculation.
The structures required to compute the energy profile are obtained by transforming the $c/a$ ratio while keeping their volumes fixed to that of the equilibrium $L1_0$ structure.
The energy profile is then evaluated by single-point calculations without geometry optimization for the structures.
Figure \ref{mlip-TiAl:Fig-bain-path} shows the energy profiles along the Bain path predicted using the EAM potential, the MLPs, and DFT calculation.
The energy profiles along the Bain path of the MLPs, particularly MLP2 and MLP3, are consistent with that obtained by DFT calculation.
The EAM profile is also close to the DFT profile for a $c/a$ ratio of 1.0--1.5, although the $c/a$ ratio of the equilibrium $L1_0$ structure of the EAM potential is slightly different from that obtained by DFT calculation.
The behavior of the energy profile along the Bain path predicted by the EAM and the deviation of the $c/a$ ratio are the same as those discussed by Zope and Mishin in Ref. \onlinecite{PhysRevB.68.024102}, who developed the EAM potential.

\subsection{Stacking faults in $\bm{\gamma}$-TiAl}

\begin{table}[tbp]
\begin{ruledtabular}
\caption{
Excessive energies of special stacking faults in $\gamma$-TiAl (unit: mJ/m$^2$).
}
\label{mlip-TiAl:table-sfe}
\begin{tabular}{cccccc}
 & EAM\footnote{Ref. \onlinecite{PhysRevB.68.024102}} & MLP1 & MLP2 & MLP3 & DFT \\
\hline
SISF $(111)$ & 108 & 281 & 165 & 258 & 194 \\
APB  $(111)$ & 249 & 680 & 616 & 681 & 694 \\
CSF  $(111)$ & 282 & 555 & 431 & 410 & 388 \\
\end{tabular}
\end{ruledtabular}
\end{table}

Computational models for the superlattice intrinsic stacking fault (SISF), the antiphase boundary (APB), and the complex stacking fault (CSF) are constructed using the procedure shown in Ref. \onlinecite{dumitraschkewitz2017impact}.
First, the supercell is constructed by the expansion along the $\langle111\rangle$ direction of $\gamma$-TiAl with ideal cubic lattice parameters, and the resultant supercell is composed of 24 atoms.
The equilibrium structure of the supercell is then obtained by local geometry optimization.
A computational model with a stacking fault is generated by introducing a tilt into the equilibrium structure.
A displacement vector defining a tilt is given by a linear combination of two vectors corresponding to the $\langle\bar110\rangle/2$ and $\langle11\bar2\rangle/2$ directions in ideal cubic $\gamma$-TiAl.
The displacement vectors for the SISF, the APB, and the CSF are expressed as
\begin{eqnarray}
\bm{b}_{\rm SISF} &=& 
\frac{1}{3} \left[ \frac{1}{2} \langle 11\bar2 \rangle \right], \nonumber \\
\bm{b}_{\rm APB} &=& 
\frac{1}{2} \left[ \frac{1}{2} \langle \bar110 \rangle \right] +
\frac{1}{2} \left[ \frac{1}{2} \langle 11\bar2 \rangle \right], \\
\bm{b}_{\rm CSF} &=& 
\frac{1}{2} \left[ \frac{1}{2} \langle \bar110 \rangle \right] + 
\frac{5}{6} \left[ \frac{1}{2} \langle 11\bar2 \rangle \right] \nonumber, 
\end{eqnarray}
respectively. 
Finally, single-point calculations are performed for the tilted structures.

Table \ref{mlip-TiAl:table-sfe} summarizes the stacking fault energies computed using the EAM potential, the MLPs, and DFT calculation.
The cohesive energy of the equilibrium structure of $\gamma$-TiAl is used as a reference to compute the stacking fault energy.
As can be seen in Table \ref{mlip-TiAl:table-sfe}, the MLPs predict the stacking fault energies of the SISF, the APB, and the CSF accurately.
On the other hand, the EAM potential lacks the predictive power for the stacking fault energy of the APB.

\begin{figure}[tbp]
\includegraphics[clip,width=\linewidth]{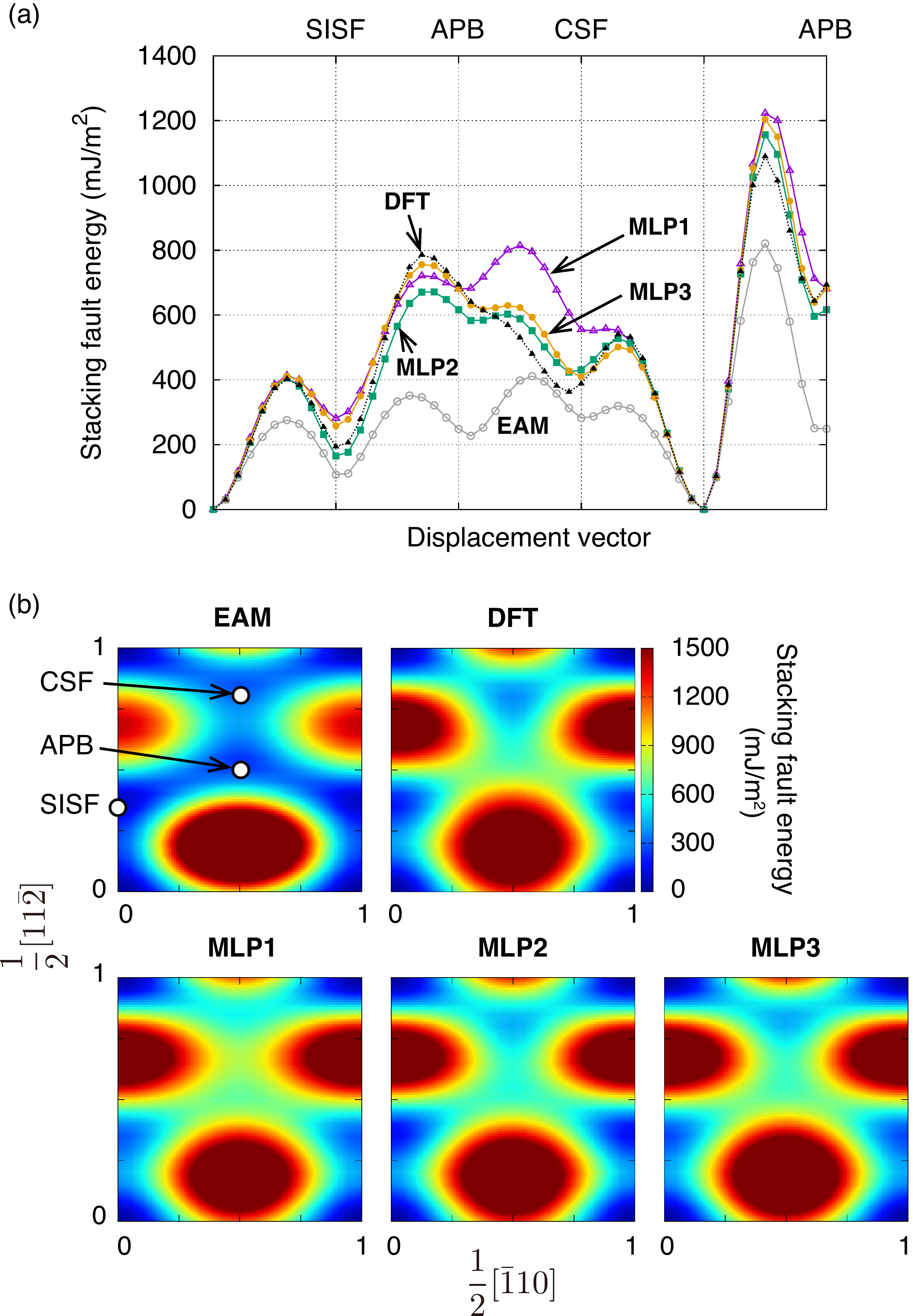}
\caption{
(a) Profile of the stacking fault energy in $\gamma$-TiAl along the path $\gamma$-TiAl $\rightarrow$ SISF $\rightarrow$ APB $\rightarrow$ CSF $\rightarrow$ $\gamma$-TiAl $\rightarrow$ APB.
(b) GSFE surface in $\gamma$-TiAl obtained using the EAM, the MLPs, and DFT calculation.
}
\label{mlip-TiAl:gsfe-map}
\end{figure}

The stacking fault energy can be defined not only for these special stacking faults but also for other general stacking faults defined by displacement vectors.
A collection of the excessive energies of the general stacking faults comprises a generalized stacking fault energy (GSFE) surface.
The displacement vector identifying a tilt of the supercell for a general stacking fault is given by
\begin{equation}
\bm{b} = u \left[ \frac{1}{2} \langle \bar110 \rangle \right] 
+ v\left[ \frac{1}{2} \langle 11\bar2 \rangle \right],
\end{equation}
where $u$ and $v$ denote the fractional coordinates defined by the vectors $\langle\bar110\rangle/2$ and $\langle11\bar2\rangle/2$, respectively.

Figure \ref{mlip-TiAl:gsfe-map} (a) shows the profiles of the stacking fault energy along the path $\gamma$-TiAl $\rightarrow$ SISF $\rightarrow$ APB $\rightarrow$ CSF $\rightarrow$ $\gamma$-TiAl $\rightarrow$ APB, predicted using the EAM potential, the MLPs, and DFT calculation.
The displacement vector changes continuously along the path.
The stacking fault energy profiles of MLP2 and MLP3 are close to that obtained by DFT calculation.
The MLP1 profile agrees with the DFT profile along a major part of the path, while it deviates from the DFT profile along the path between the APB and the CSF.
The EAM profile also deviates from the DFT profile along the path between the SISF and the CSF.
Figure \ref{mlip-TiAl:gsfe-map} (b) shows the GSFE surface in $\gamma$-TiAl predicted using the EAM potential, the MLPs, and DFT calculation.
The GSFE surfaces predicted using the MLPs and DFT calculation are similar, while that predicted using the EAM potential is different from that predicted by the DFT calculation.
Thus, the present MLPs should have high predictive power for the stacking faults and related properties.

\section{Conclusion}
\label{mlip-TiAl:Sec-Conclusion}

In this study, O(3) polynomial invariants representing the neighboring atomic density in a multicomponent system have been formulated.
Polynomial models have also been introduced to describe the relationship between the atomic energy and the polynomial invariants. 
Although the present formulation to develop MLPs in a multicomponent system is more complex than the formulation for elemental systems, it enables the accuracy and computational efficiency of MLPs to be controlled systematically.

This study also shows an application of the formulation of MLPs to the Ti-Al binary alloy system.
Pareto optimal MLPs have been developed by applying the polynomial models combined with the polynomial invariants.
These MLPs are available in the \textsc{Machine Learning Potential Repository} \cite{MachineLearningPotentialRepositoryArxiv,MachineLearningPotentialRepository}.
The predictive power of the Pareto optimal MLPs has been examined for the cohesive energy, the formation energy, the elastic constants, the phonon density of states and dispersion curves, the thermal expansion, the energy profile along the Bain path, and the stacking fault properties. 
The MLP with the lowest prediction error (MLP3) and that with high computational cost performance (MLP2) have high predictive power for all the properties, whereas an MLP showing higher computational cost performance (MLP1) than MLP2 fails to predict some of the properties with high accuracy. 
This study reveals that the present framework provides a systematic way to develop MLPs with high computational cost performance in multicomponent systems.

\begin{acknowledgments}
This work was supported by a Grant-in-Aid for Scientific Research (B) (Grant Number 19H02419), and a Grant-in-Aid for Scientific Research on Innovative Areas (Grant Number 19H05787) from the Japan Society for the Promotion of Science (JSPS).
\end{acknowledgments}

\bibliography{mlip-TiAl}
\end{document}